\documentclass[useAMS,usenatbib]{mnras}
\usepackage{graphicx}
\usepackage{graphics}
\usepackage{multicol}
\usepackage{etex}
\usepackage{setspace}
\usepackage{float}
\usepackage{pdflscape}
\usepackage{multirow}
\usepackage{hhline,booktabs,amsmath}
\usepackage{makecell}
\usepackage{caption}
\usepackage{subcaption}

\usepackage{ae,aecompl}

\usepackage{graphicx}
\usepackage{rotating}
\usepackage{epsfig}
\usepackage{amsmath}
\usepackage[utf8]{inputenc}
\usepackage{array}
\usepackage{ragged2e}
\usepackage{longtable}
\usepackage{supertabular}
\usepackage{subfloat}
\usepackage{graphicx}	
\usepackage{amsmath}	

\usepackage{newtxtext,newtxmath}

\title{Exploring the inner-disk region of the atoll source 4U 1705-44 using AstroSat's SXT and LAXPC observations}

\author[S. Malu et al.]{
Malu S,$^{1}$\thanks{E-mail:malu.sudhaj@gmail.com}
K. Sriram,$^{1}$
S. Harikrishna,$^{1}$
Vivek K. Agrawal$^{2}$
\\
$^{1}$Department of Astronomy, University College of Science, Osmania University, Hyderabad, India\\
$^{2}$Space Astronomy Group, ISITE Campus, U R Rao Satellite Center, 560037, Bangalore, India \\
}

\date{Accepted XXX. Received YYY; in original form ZZZ}

\pubyear{2021}

\begin{document}
\label{firstpage}
\pagerange{\pageref{firstpage}--\pageref{lastpage}}
\maketitle
﻿

\begin{abstract}
For the first time, simultaneous broadband spectral and timing study of the atoll source 4U 1705-44 was performed using AstroSat Soft X-ray Telescope (SXT) 
and Large Area X-ray Proportional Counter (LAXPC) data (0.8-70 keV). Based on the HID, the source was in the soft banana state during these observations. Spectral modeling was performed using the full reflection framework and an inner disk radii of 14 Rg was obtained.  A hard powerlaw tail was noticed in the soft state and hot component fluxes and varying powerlaw indices point towards a varying corona/sub-keplerian flow. 
Based on the spectral fits the boundary layer radius and magnetospheric radius were constrained to be $\sim$ 14-18 km and $\sim$ 9-19 km respectively. 
Cross-Correlation Function studies were performed between the 0.8-3 keV soft SXT lightcurve and 10-20 keV hard LAXPC lightcurve and correlated and anticorrelated lags were found, 
which was used to constrain the coronal height to 0.6-20 km ($\beta$=0.1). 
Since the inner disk radius is not varying during the observations, we conclude that the detected lags are possibly caused by a varying structure of corona/boundary layer in the inner region of the accretion disk. Based on the observations,
a geometrical model is proposed for explaining the detected lags in the atoll source 4U 1705-44.

\end{abstract}




\section{Introduction}

The geometry and structural configuration of Neutron Star Low mass X-ray binaries (NS LMXBs)
is an open question, especially with respect to the location and geometry of the Corona. 
Spectral and timing differences have been noted for the two major subcategories of NS LMXBs,
namely, Z and atoll sources, but a unified model, especially to address the geometry, is currently lacking.
Z sources trace out a Z track in Hardness Intensity
Diagram (HID) and atoll sources trace out a C shaped track in the HID with Z sources being highly luminous sources 
in the X-ray regime with L $\sim$ 0.5-1 L$_{Edd}$ (van der Klis 2005), while atoll sources are less luminous X-ray 
sources with L $\sim$ 0.001-0.5 L$_{Edd}$ (eg. Ford et al. 2000, van der Klis 2006). 
Atoll sources have two primary branches called island and banana states (Hasinger \& van der Klis 1989; 
van der Klis 2006), with the banana state being further categorized into the lower and upper banana state. 
The banana state is the soft state with higher luminosities while the island state is characterized by harder spectra and lower luminosity (Barret 2001; Church et al. 2014). The causative mechanism that drives these sources along the HID is an ambiguous territory with some claims of it being due to varying mass accretion
rates and others being instabilities due to accretion flow solutions or radiation pressure at the inner disk
radius or boundary layer around the NS surface (Homan et al. 2002, 2010; Lin et al. 2007).

The picture understood so far has been that of a disk formed via Roche lobe overflow from the secondary star,
with the disk being truncated (Esin et al. 1997; see Done et al. 2007), a boundary layer around the NS surface 
which is the site where the disk material reaches the relatively slowly spinning NS (Shakura \& Sunyaev 1988; 
Popham \& Sunyaev 2001), a compact corona/hot electron cloud close to the NS (Sunyaev et al. 1991). In the island state of atoll sources the disk has been considered to be 
comparatively further away from the NS as opposed to the banana state where it approaches the ISCO (Barret \& Olive 2002, 
Egron et al. 2013). But the exact location or structural setup in this region or the extent
of the corona and its nature is not clearly understood. A combination of broadband spectral and timing studies, 
specifically time lag studies (Cross Correlation Function) may contribute towards this cause by providing an
opportunity to utilize information about the different soft and hard X-ray emitting regions.

Spectral model degeneracy is an issue when it comes to these sources but the spectrum is considered to be mainly modeled using two
components viz. a soft/thermal component and a hard/Comptonized component. While the classic Eastern model (Mitsuda et al. 1984)
uses a multicolor disk blackbody (MCD) for the thermal component and a weakly Comptonized blackbody for the Comptonzed component,
the Western model (White et al. 1988) uses a single temperature blackbody (BB) for the emission from the boundary layer and a
Comptonized emission from the disk. Lin et al. (2007) used a hybrid model based on the study of two atoll sources, that uses
a single temperature blackbody and a broken powerlaw (BPL) when the source is in the hard state and a constrained BPL and
two thermal components (MCD and BB) in the soft state. This model presents a weak Comptonization solution that differs as opposed
to the strong Comptonization solution offered by the two component models proposed prior to this. Cackett et al. (2010) used a 
power-law instead of a BPL for the same hybrid model when modeling the continuum for a sample consisting of both atoll
and Z sources. The hard X-ray photons originating from the boundary layer or the Corona illuminates the disk and the disk material reprocesses it, producing a reflection spectrum.  
Reflection modeling in NS LMXBs can help constrain the radial extent of NS (Cackett et al.2008; Miller et al. 2013).
Still, a consensus on a model that best describes the inner accretion disk corona geometry is lacking. More broadband spectral studies encompassing the extremely soft and hard regimes
of such NS LMXBs can help in gaining a better understanding of the state of these sources.

The timing technique of Cross-Correlation Function (CCF) study between soft and hard X-ray energy bands can be used to probe the regions emitting these photons, as the relation between the emission of
soft and hard photons are associated with the relative variations in the structure or configuration
of the soft and hard X-ray emitting regions. Lags obtained between soft and hard energy X-ray photons
vary from short to long timescales. Lei et al. (2013) performed CCF studies on the atoll source 4U 1735-44
and found soft and hard lags of a few hundred seconds, where soft lag refers to the lagging of soft X-ray photons to
hard photons and similarly hard photons lagging to soft photons are termed as hard lags. 
They interpreted the anticorrelated lags to be due to
variation in the disk structure. Sriram et al. (2012) performed CCF studies on the Z source GX 5-1 and
found few ten to hundred seconds lag in the HB and NB of the HID, which were explained in the context
of the re-condensation of the coronal material within the inner region of the accretion disk.

CCF studies performed by Sriram et al. (2019) on the Z source GX 17+2, using RXTE and NuStar satellite data, showed evidence of the CCF lags being caused
primarily by the readjustment of the corona. In this study, the disk was found to be almost at the last stable orbit and the viscous time scale
of the disk was estimated to be few tens of seconds. This would imply that the few hundred seconds lag should be the readjustment time scale of the coronal structure, and an equation was arrived at for estimating coronal height based on the observed lags. 
Here the coronal velocity was assumed to be a $\beta$ times that of the disk velocity where $\beta$ $\le$ 1. Coronal height
was constrained to be of the order of few tens of km. Similarly for the first time, using AstroSat LAXPC (hard) and SXT (soft) data of GX 17+2, 
Malu et al. (2020) found correlated and anticorrelated lags of the order of a hundred seconds, which were used to constrain
the coronal height to few tens of km. AstroSat LAXPC hard and soft lightcurves were used by Sriram et al. (2021) to perform CCF studies
which again revealed a few ten to hundred seconds lag in the HB and NB of GX 17+2. Such a study based on longer CCF lags can reveal vital information about the physical variability of the inner region of the disk and the coronal structure.

A few milli-seconds hard time lag is indicative of the Comptonization process which reprocesses the soft photons
from the disk or the NS surface to hard photons in the hotter Compton cloud/jet (Vaughan et al. 1999; Kotov et al. 2001; Qu et al.
2004; Arevalo \& Uttley 2006; Reig \& Kylafis 2016). In the opposite scenario
of a few milli-second soft lag (soft photons lagging the hard photons, negative lag in the CCF profile), a shot model
(Alpar \& Shaham 1985) or a two-layer Comptonization model (Nobili et al. 2000) have been used.
Similar timing studies of atoll sources may help in better understanding the coronal/sub-keplerian flow and structure around the NS.

4U 1705-44 is a type 1 X-ray burster, that is an atoll type NS LMXB (Hasinger \& van der Klis 1989). It is located
at a distance of 7.4 kpc (Haberl \& Titarchuk 1995, Galloway et al. 2008). Piraino et al. (2007) found an inclination angle of 20$^\circ$ - 50$^\circ$
based on the spectral fitting of BeppoSAX observations of the source, whereas the spectral modeling of Chandra observations yielded
an inclination angle of 58$^\circ$ - 84$^\circ$ (Di Salvo et al. 2005). The Fe emission line has been observed in the hard and soft states.
Di Salvo et al. (2015) studied the source in the hard state and found reflection features
in the spectrum with parameters consistent with that in the soft state with an inner disk radii that is not truncated at a larger radius. D' Ai et al. (2010) modeled the hard state with a very soft thermal disk emission with a larger inner disk radius
and a thermal Comptonized emission. The iron line was also explained using the reflection model even with a truncated accretion disk.
While Di Salvo et al. (2009) arrived at an inner disk radius (R$_{in}$) of $\sim$ 2.3 ISCO based on the spectral modeling of the iron emission line
in the XMM-Newton observations, Reis et al. (2009) obtained an R$_{in}$ value of 1.75 ISCO based on Suzaku observations. Phenomenological modeling of the asymmetric Fe K emission line, by Cackett et al. (2010), constrained R$_{in}$ to be 1.0--6.5 ISCO.
Egron et al. (2013) found that for 4U 1705-44 in the soft state R$_{in}$ $\sim$ 10--16 Rg and up to 26--65 Rg in the hard state.
AstroSat LAXPC observation of the source  when it was in the soft banana state was modeled by Agrawal et al. (2018) which resulted in an R$_{in}$ of 26--68 km. This was based on the DiskBB normalization parameter used in the continuum modeling which was then also colour corrected.  A hard X-ray tail has been noted when the source is in the soft state (Piraino 
et al. 2007, 2016, Lin et al. 2010) using BeppoSAX observations and Suzaku observations. Using LAXPC observations of the source. Agrawal et al. (2018)
has also noted a varying powerlaw tail in the soft state. 

For the first time using simultaneous AstroSat SXT and LAXPC data, the energy dependent CCF studies between the soft (0.8-3 keV) and hard (10-20 keV)
is being reported here and a geometrical model to explain the lags is presented. A broadband spectral study (0.8-70 keV) has been performed for the source
using simultaneous SXT and LAXPC spectral data which has not been previously done. The soft energy band information ( <3 keV) in both temporal and spectral analysis
can be very crucial to get a complete picture of the geometrical configuration of the source, especially the truncation radius.

\section{Data Reduction and Analysis}

Using AstroSat, regular pointing observations of 4U 1705-44 were made for 12 satellite orbits based on our proposal in AO cycle 3 
(Obs ID: A03\_073T01\_9000001498) from 2017, August 29, 01:53:37 to 2017, August 30, 01:06:48
for an effective exposure time of $\sim$ 36.8 ks.
We have used both LAXPC (Large Area X-ray Proportional Counter) and SXT (Soft X-ray Telescope) simultaneous
data of the source. LAXPC comprises of 3 identical proportional counter units (LAXPC 10,20 and 30) with a total
effective area $\sim$ 6000 cm$^2$ at 15 keV and it operates in the 3-80 keV energy range 
(Yadav et al. 2016a; Agrawal et al. 2017; Antia et al. 2017). We have used data in the Event Analysis (EA) mode
with a time resolution of 10$\mu$s. SXT operated in the 0.3-8 keV energy range and has an effective area of $\sim$ 128 cm$^2$ at 1.5 keV (Singh
et al. 2017). We have used data in the Photon Counting (PC) mode which has a time resolution of $\sim$ 2.4 s.

LAXPC level 1 data were processed using the LAXPC software (Format A, May 19, 2018 version)\footnote{http://astrosat-ssc.iucaa.in/?q=laxpcData}
provided by the AstroSat Science Support Center (ASSC). Standard procedures for data reduction were followed, with event files and
GTI files were generated using the laxpc\_make\_event and laxpc\_make\_stdgti routines. Lightcurves and spectra were generated using the laxpc\_make\_lightcurve and laxpc\_make\_spectra routines. Corresponding standard routines were used to generate the background files.
laxpc\_make\_spectra routine also generates the appropriate response files needed for the spectra.

A merged cleaned event file was generated for the SXT level 2 data using the event merger code and the appropriate ancillary response file was generated
using the sxtARFmodule, both provided by the SXT team \footnote{https://www.tifr.res.in/$\sim$astrosat\_sxt/dataanalysis.html}. Lightcurves and spectra
were extracted from the SXT image of the source using XSELECT (V2.4e). For spectral analysis, response file (sxt\_pc\_mat\_g0to12.rmf) and deep blank sky background spectrum
file (SkyBkg\_comb\_EL3p5\_Cl\_Rd16p0\_v01.pha) provided by the SXT team were utilized. Similar to the procedure adopted in Malu et al. (2020), 
the source was extracted with a 1$\arcmin$-10$\arcmin$ annulus to take care of the pile-up effect (see Figure 1), since the count rate was  $>$ 40 cts s$^{-1}$  (see AstroSat handbook \footnote{http://www.iucaa.in/$\sim$astrosat/AstroSat\_handbook.pdf}).
Background extraction for the lightcurves was performed using a 13$\arcmin$-15$\arcmin$ annulus. See Figure 2 for the extracted SXT and LAXPC lightcurves.

\begin{figure}
\includegraphics[height=6cm,width=8cm, angle=0]{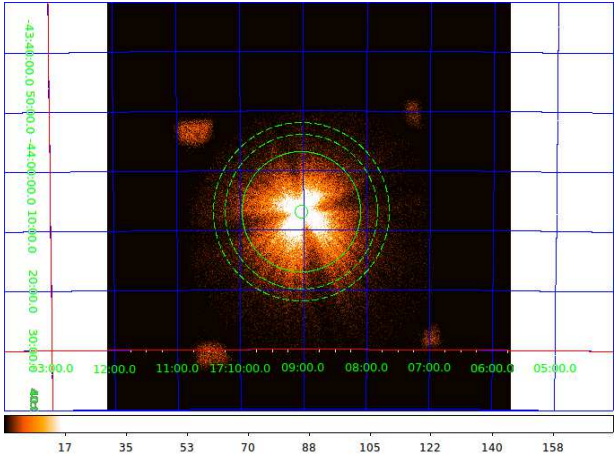}\\

\caption{ Top: SXT image of the source and the extraction regions used for source (solid annulus) and background (dashed annulus) extraction. Calibration
sources can be seen in the four corners of the image.}
\end{figure}

\begin{figure}
\includegraphics[height=8cm,width=6cm, angle=270]{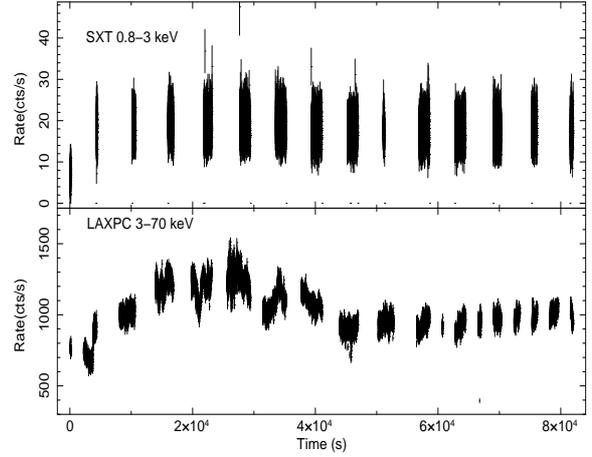}\\

\caption{ Top: SXT light curve in the 0.8–3 keV energy range. Bottom:
LAXPC 10 light curve in the 3–70 keV energy range. Both lightcurves are plotted using a time bin of SXT resolution $\sim$ 2.4 s}
\end{figure}

LAXPC 10 data alone was used for analysis as it is well-calibrated and has less background issues (Agrawal et al. 2020).
For spectral analysis, LAXPC 10 top layer spectra alone were used to minimize the background (eg. Beri et al. 2019).  
Joint spectral analysis were performed using the LAXPC spectrum in the range 5-70 keV and SXT spectrum in the range 0.8-7 keV. 
Owing to uncertainties in response, data below 0.8 keV were not considered for SXT (Bhargava et al. 2019). 
 A 3\% systematic error was introduced while performing spectral fitting \footnote{https://www.tifr.res.in/$\sim$astrosat\_sxt/dataana\_up/\\readme\_sxt\_arf\_data\_analysis.txt\label{note4}} 
 (eg. Jithesh et al. 2019, Bhargava et al. 2019).

\section{Spectral analysis}   

HID was obtained using the 3.0-18.0 keV, 7.5-10.5 keV, and 10.5-18.0 keV energy bands (similar to Agrawal et al. 2018) (Figure 3). Hardness ratio was
taken as the ratio between 7.5-10.5 keV and 10.5-18.0 keV and intensity was that in the range of 3.0-18.0 keV. 
HID revealed the source to be in the banana state and the branch was divided into 3 sections - A, B, and C for performing
spectral analysis. 
For each of the three sections of the HID viz. A, B and C, spectral modeling was performed using XSPEC v12.10.0c (Arnaud 1996).
Only those portions that are observed simultaneously with LAXPC and SXT are used for performing a joint spectral fit (Figure 4).
SXT 0.8-7 keV and LAXPC 5-70 keV was used for producing the broadband 0.8-70 keV joint spectra.
Hydrogen column density N$_H$, modeled with the {\it Tbabs} (Wilms et al. 2000), was allowed to vary for A section which was then fixed for the other two sections. To model the joint spectra, a constant factor was also involved in the fit to take into
account the relative normalization factor SXT and LAXPC spectra. While performing the fit, the gain correction was made for SXT data using
the gain fit command, where the slope was fixed at unity and offset was freed \textsuperscript{\ref{note4}}.

\begin{figure}
\includegraphics[height=8cm,width=6cm, angle=270]{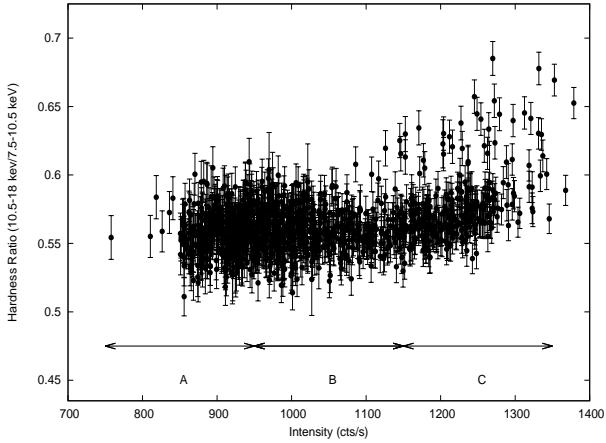}\\

\caption{ HID for 4U 1705-44 using AstroSat LAXPC observations. Hardness ratio is defined as
10.5-18.0/7.5-10.5 keV and intensity is that in the range 3.0-18.0 keV range. Demarcations show separations of HID into
A, B and C sections.}
\end{figure}

\begin{figure}
    \includegraphics[height=8cm, width=\linewidth,angle=0]{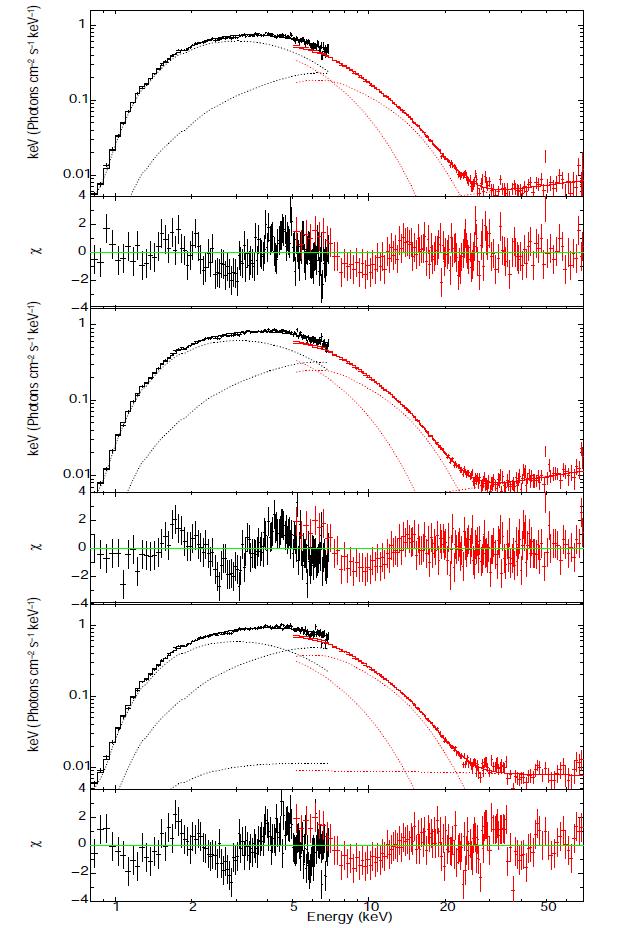}
\caption{The SXT+LAXPC joint spectral fit in 0.8-70 keV for the the {\it Diskbb + rdblur*bbrefl + Power-law} model for sections A, B and C (top to bottom) respectively. The top panel gives
the unfolded spectra (thick line) with the component models (dashed lines) and the respective bottom panel gives the residuals obtained from the fit.}
\end{figure}

We initially modeled the spectral continuum using a {\it DiskBB+Bbody+Powerlaw} model following Cackett et al. (2008, 2010) and Lin et al. (2007).
This resulted in $\chi^2$/dof values of 421.4/354, 528.3/355, 550.1/355 for A, B, and C respectively. A residual around 6.4 keV was noticed which was modeled using a Gaussian function with the centroid energy fixed at 6.4 keV. Its line width and normalization were allowed to vary. This fit resulted in a $\chi^2$/dof value of 
323/352, 357.5/353, and 381.6/353 for A, B, and C sections respectively. We note a hard tail at energies above 25 keV in each of the spectra, 
similar to Piraino et al. (2007), Agrawal et al. (2018).  
Although the fit resulted in good $\chi^2$/dof values, the powerlaw component could not
be effectively constrained, resulting in high error bars (see Table 1). Blackbody temperature KT$_{bb}$ was found to be 2.63$_{-0.15}^{+0.21}$ keV in section A, 3.03$_{-0.25}^{+0.35}$ keV in section B
and dropped to 1.16$_{-0.08}^{+0.09}$ keV in section C, while DiskBB temperature was found to be 1.95$_{-0.08}^{+0.09}$ keV (A), 2.25$_{-0.08}^{+0.07}$ keV (B) and a higher 2.60$_{-0.03}^{+0.02}$ keV (C).

As inferred from the DiskBB normalization, the disk is relatively close to the NS, and hence the reflection effect would play an important
role. A relativistic blurring kernel has been used to convolve this reflection spectrum. This will take into account the effects on the reprocessed
emission from the disk. Hence for a more physically realistic picture, we used a blurred reflection model to fit the spectra (Figure 4).
In the full reflection framework, {\it bbrefl} (Ballantyne 2004) was utilized to model the reflection from the disk by a blackbody, and it was relativistically blurred with the convolution model {\it rdblur} (Fabian et al. 1989) (eg. Mondal et al. 2016, Cackett 2016, Malu et al. 2020) (see Table 2).
Here an iron abundance of 1 was used (Malu et al. 2020).
Along with this, {\it DiskBB+Powerlaw} models were also used. Emissivity index in the {\it rdblur} model was fixed to -3 in all the spectra and
the outer radius was fixed to 1000 Rg (Rg=GM/c$^{2}$). Based on Di Salvo et al. (2005), the inclination was fixed to an average value of 70 $^{\circ}$.
The inner disk radius was found to be $\sim$ 14 Rg ($\sim$ 29 km) within error bars in all three sections, indicating a stationary disk. 
This value is very well in agreement with that obtained from previous studies by Di Salvo et al. (2009), D' Ai et al. (2010), and Egron et al. (2013).
The reflection fraction was constrained to be 1.63 $_{-0.08}^{+0.08}$ (A), 1.63 $_{-0.06}^{+0.06}$ (B) and 1.64 $_{-0.05}^{+0.05}$ (C). 
The R$_{in}$ obtained from {\it DiskBB} model after color correction and {\it rdblur} model are matching (within error bars, see Table 2). 
The ionization parameter was found to be log$\xi$ 3.42 $_{-0.12}^{+0.12}$ (A), 3.41 $_{-0.13}^{+0.11}$ (B) and 3.45 $_{-0.11}^{+0.11}$ (C). 
The incident blackbody temperature varied from $\sim$ 2.21$_{-0.01}^{+0.01}$ keV (A)--2.04$_{-0.05}^{+0.05}$ keV (C) showing a slight decrease along the branch, while the Diskbb temperature
KT$_{in}$ remained the same within error bars along the branch (See table 2). Previously, using LAXPC data, Agrawal et al. (2018) found the powerlaw index
pertaining to the hard tail to have $\Gamma_{pl}$ $\sim$ 0.79--3.41, which contributed to 4-30 \% of the total flux. Our obtained values are closely
in agreement with their values, with $\Gamma_{pl}$ 0.59$_{-0.21}^{+0.22}$ (A), 0.48$_{-0.17}^{+0.16}$ (B) and 1.07$_{-0.18}^{+0.19}$ (C), which
contributed to 5.85 \% - 7.12 \% of the total flux. There was a softening of the powerlaw index from A to C, 
and the powerlaw flux is found to increase slightly as the source moved from A to C along with the blackbody flux which was also found to be increasing along the branch (see Table 2).
This fit resulted in $\chi^2$/dof values of 326.5/353, 357/354, and 395/354 for sections A, B, and C respectively.
R$_{in}$ from the {\it rdblur} model was further constrained by computing $\Delta$$\chi$$^2$ (=$\chi^2$-$\chi^2_{min}$) using the steppar command in xspec for each of the parameters in the best fit model (Figure 5).

 Apart from this model, we also tried a {\it Nthcomp (hard component)+Bbody (soft component) +Gaussian+Powerlaw} model which resulted in  $\chi^2$/dof values of 274/350, 270/351, and 327/351
respectively for A, B, and C sections (Table 3). Seed photon temperature KT$_{bb}$ was found to be 0.96$_{-0.06}^{+0.06}$ keV in section A, 1.05$_{-0.04}^{+0.04}$ keV in section B
and 1.07$_{-0.05}^{+0.05}$ keV in section C, while the electron temperature was found to be 2.92$_{-0.13}^{+0.17}$ keV (A), 2.96$_{-0.18}^{+0.22}$ keV (B) and 
2.79$_{-0.16}^{+0.20}$ keV (C).  $\Gamma_{pl}$ was noted to be 0.36$_{-0.27}^{+0.25}$ (A), 0.13$_{-0.22}^{+0.20}$ (B) and 0.63$_{-0.28}^{+0.26}$ (C). No definite trend of increase or decrease in the spectral parameters of this model was observed here.

\begin{figure}
\includegraphics[width=4.0cm,height=8.0cm,angle=270]{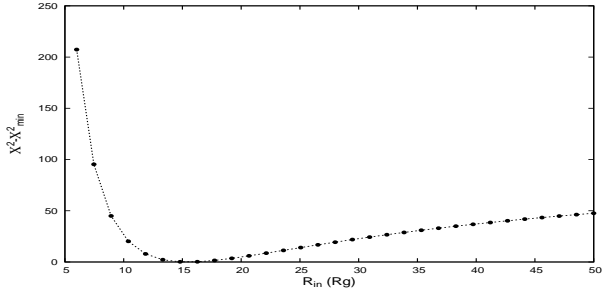} \\
\caption{$\Delta$ $\chi^2$ (=$\chi^2$-$\chi^2$$_{min}$) vs R$_{in}$ for the best fit model (Table 2).}
\end{figure}

\section{Timing analysis}

LAXPC 10 and SXT (bin size of 2.3775 s i. e. SXT time resolution) lightcurves of 4U 1705-44 are shown in Figure 2. 
Cross Correlation Function (CCF) studies were performed between soft (0.8-3 keV) and hard (10-20 keV) X-ray light curves from LAXPC and SXT. 
The {\it crosscor} tool available in the XRONOS package was utilized for performing the CCF analysis (eg. Sriram et al. 2007, 2011a; Lei et al. 2008, Malu
et al. 2020). A direct slow algorithm was employed to perform the CCF analysis. In this mode, the theoretical error bars of the cross correlation
from individual intervals are propagated to obtain the CCF error bars \footnote{https://heasarc.gsfc.nasa.gov/docs/xanadu/xronos/help/crosscor.html}.
 
A 20 s bin size was used for the procedure and CCFs were obtained between the simultaneous 0.8-3 keV SXT and 10-20 keV LAXPC lightcurves (Figure 6a-c).
Lags were obtained in two different light curve segments, out of which one is a correlated soft lag while the other is an anticorrelated hard lag.
Here soft lag means that the soft photons (0.8-3 keV) are lagging behind the hard (10-20 keV) photons and vice versa for hard lag.
To estimate the lag values Gaussian functions along with a constant factor were fitted around the most significant peak
(highest CC) in the CCF profile. The fit was performed with a 90 \% confidence level using a $\chi^2$ minimization method with the criterion of 
$\Delta$$\chi$$^2$ = 2.7  (Figure 7a). For the asymmetric CCF profile, we used two Gaussian functions with a constant factor for obtaining the lag.
Using this fit procedure we obtained a soft lag of -38 $\pm$ 18 s in one lightcurve segment with a CC value of 0.67 $\pm$ 0.27, while another segment
exhibited an anticorrelated hard lag of 259 $\pm$ 47 s with a CC value of -0.39 $\pm$ 0.16. Another segment showed a positive correlation with no lag with
a CC value of 0.47 $\pm$ 0.15. The remaining segments exhibited uncorrelated CCFs with CC $<$ 0.3.

\begin{figure}
  \subcaptionbox{ }[0.8\linewidth][c]{%
    \includegraphics[height=10cm, width=0.7\linewidth,angle=270]{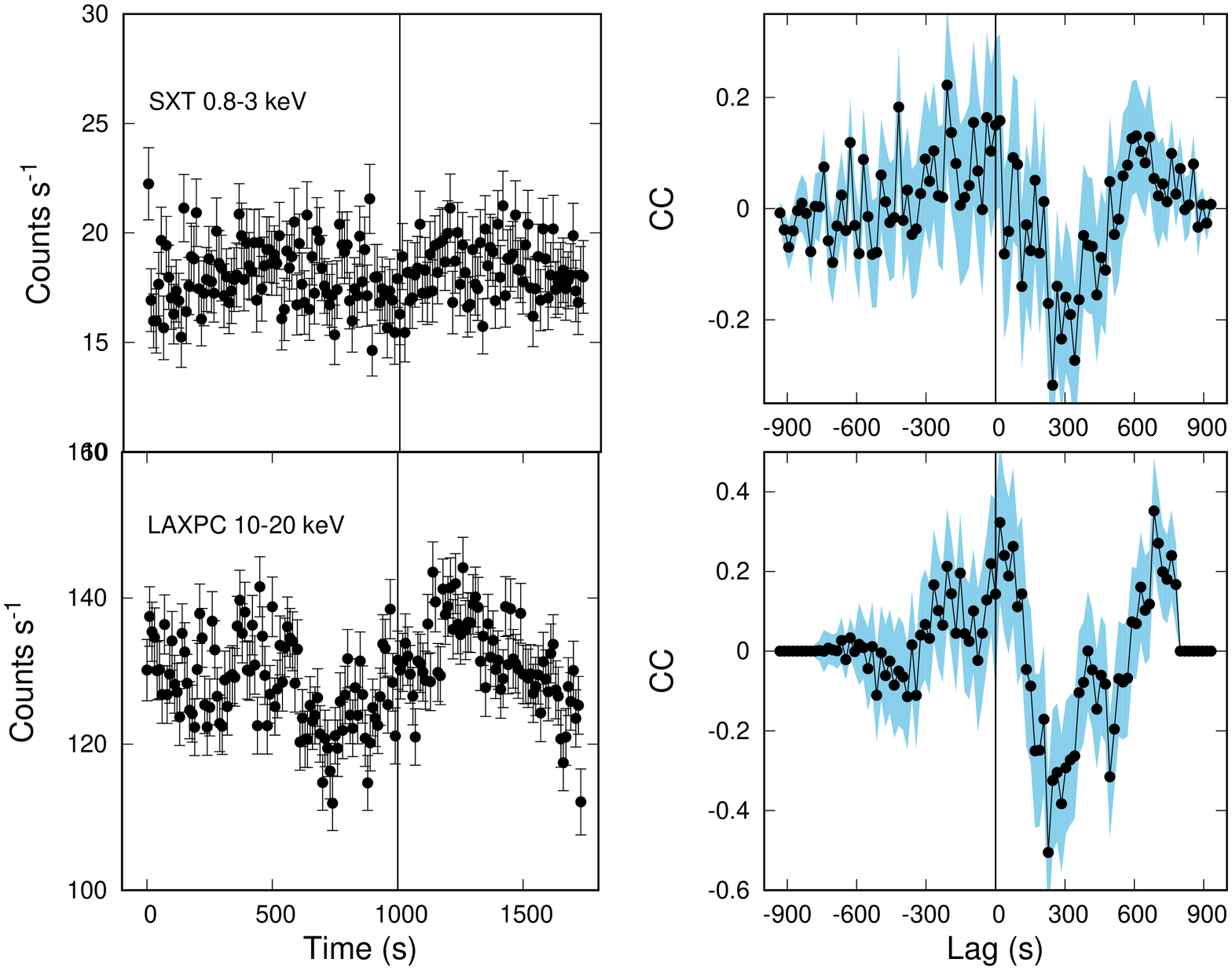}}\quad
  \subcaptionbox{ }[0.8\linewidth][c]{%
    \includegraphics[height=10cm, width=0.7\linewidth,angle=270]{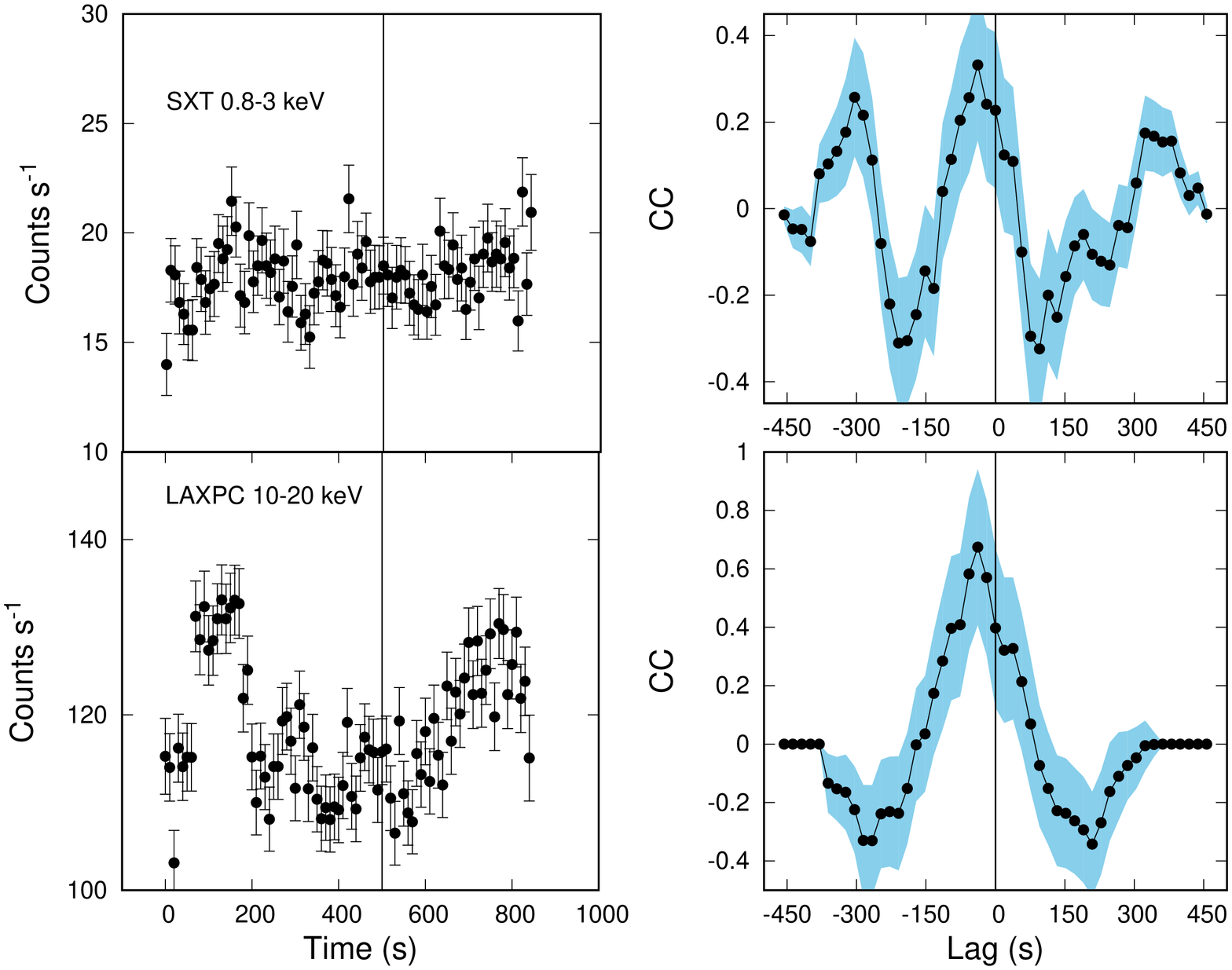}}\quad
  \subcaptionbox{ }[0.8\linewidth][c]{%
    \includegraphics[height=10cm, width=0.7\linewidth,angle=270]{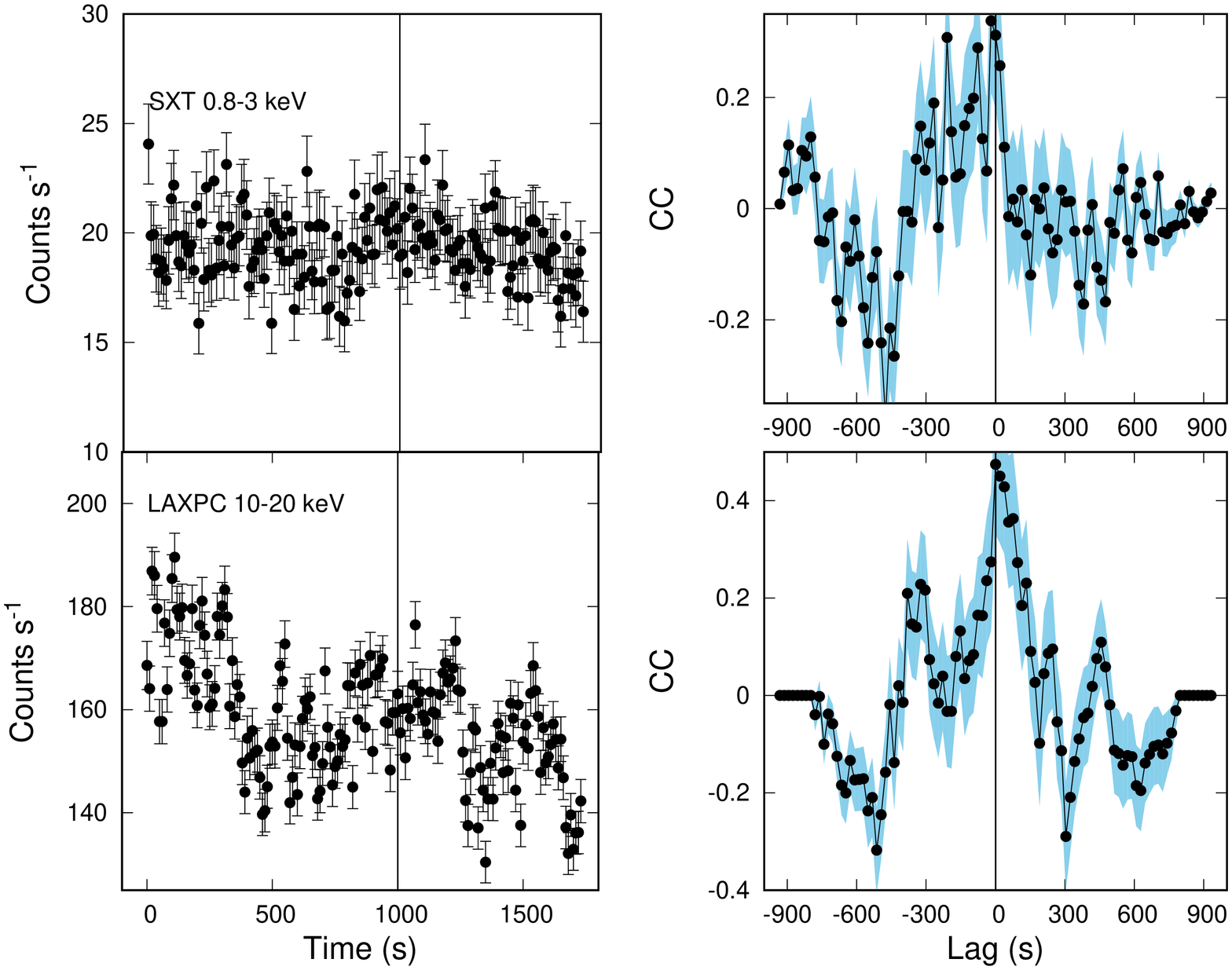}} 
\caption{The background subtracted SXT soft (0.8-3 keV) and LAXPC hard X-ray (10--20 keV) light curve (left panels) for which CCF lag is observed (right panels).
Energy bands used are mentioned in the light curves (left panel). Right panels show the cross correlation function (CCF) of each section of the light curve and shaded regions show the standard deviation 
of the CCFs.}
\end{figure}

\begin{figure}
  \subcaptionbox{ }[0.8\linewidth][c]{%
    \includegraphics[height=9cm, width=0.5\linewidth,angle=270]{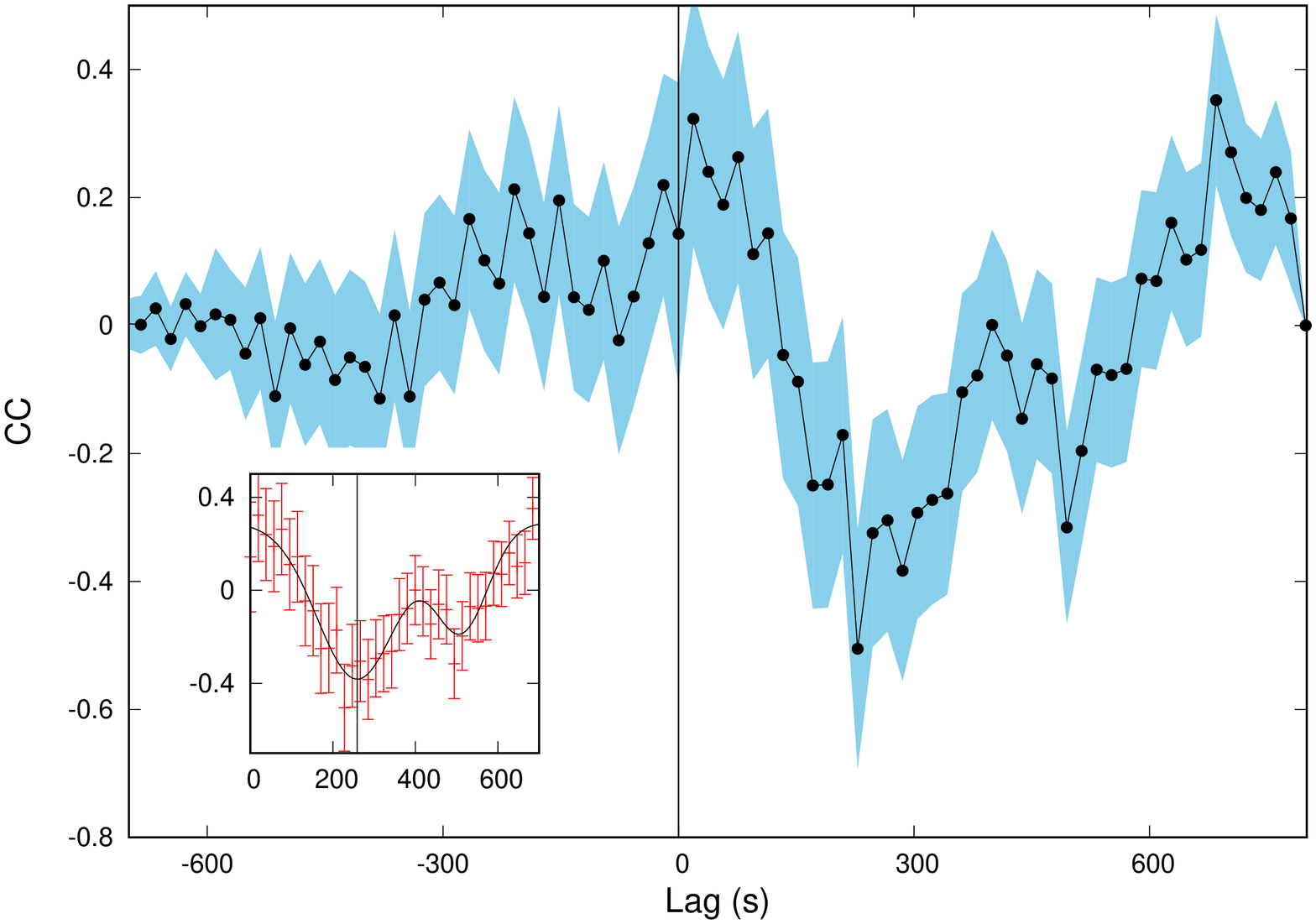}}\quad
  \subcaptionbox{ }[0.8\linewidth][c]{%
    \includegraphics[height=9cm, width=0.5\linewidth,angle=270]{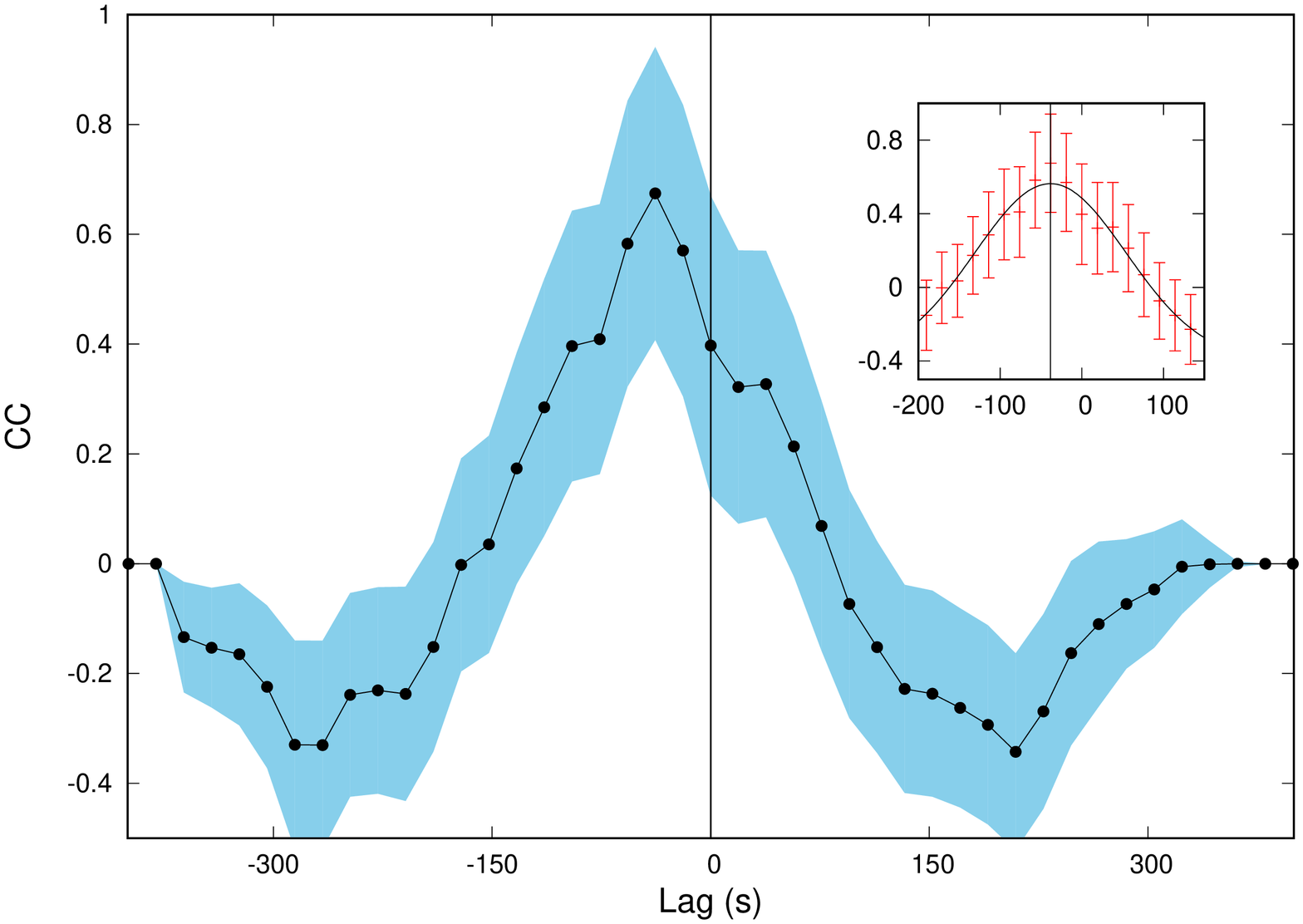}}
\caption{Figure shows CCF lags fitted using a Gaussian function using points around the maximum peak.}
\end{figure}

\section{Results and Discussion}

\subsection{Inner disk radius and Boundary layer}

The full reflection framework used for spectral modeling employed a {\it Diskbb + rdblur*bbrefl + power-law}  model.
Diskbb model normalization (Mitsuda et al. 1984) is given by N = (R$_{in}$/D$_{10}$)$^{2}$ $\times$ cos i, where D$_{10}$ is
the distance to the source in units of 10 kpc and  R$_{in}$ is the inner disk radius. Using an inclination angle (i) of 70$^{\circ}$,
we estimated R$_{in}$ to be 8.49 km, 8.25 km, 8.54 km. These values are apparent inner disk radii and for estimating the true radii
correction factors for spectral hardening ($\kappa$) and inner boundary condition ($\xi$) were used. Shimura \& Takahara (1995) gave
a $\kappa$ value $\sim$ 1.7--2.0, and Kubota et al. (1998) gave a $\xi$ value of 0.41. Hence,
R$_{eff}$ can be given as $\kappa^{2}$ $\xi$ R$_{in}$ (Kubota et al. 2001). Our estimated R$_{eff}$ values are $\sim$
10-14 km (A), 10-13 km (B) and 10-14 km (C). Hence the disk appears to be fairly close to the NS and
also stationary along the banana branch. These values are in agreement with Di Salvo et al. (2005), where R$_{in}$ was noted to be 15 km.

The convolution model {\it rdblur} gives an inner radii of 14.99$_{-3.49}^{+8.02}$ R$_{g}$, 14.45$_{-2.69}^{+5.32}$ R$_{g}$ and 13.57$_{-2.52}^{+4.52}$ R$_{g}$, which is
$\sim$ 31.10 km, 29.98 km and 28.15 km respectively for A, B and C sections which also supports the stationary inner disk inference. 

The boundary layer (BL) between the inner disk and the NS is a major source of the hot blackbody or the Comptonized emission
that dominates around 7--20 keV (e.g. Popham \& Sunyaev 2001, Barret et al. 2000) and can act as a significant contributor
to the accretion luminosity. The idea of a Comptonized emission spectrum from a boundary layer was supported by the Fourier
frequency resolved spectrum (Gilfanov, Revnivtsev \& Molkov 2003, Revnivtsev \& Gilfanov 2006).

Hence we estimate the radius of this BL using the equation given by Popham \& Sunyaev (2001),
\begin{equation}
log(R_{BL} - R_{NS}) \sim 5.02 + 0.245 \Bigg[log\Big({\frac{\dot{M}} {10^{-9.85}\ \ M_{\odot} yr^{-1}}\Big)}\Bigg]^{2.19}
\end{equation}

This value for sections A,B and C came out to be, 15.74 km (A and B) and 16.52 km (C) respectively using the mass accretion rate obtained from 
the equation L= $\frac{GM\dot{M}}{R}$ with M = 1.4 M$_{\odot}$, R = 10 km and luminosity from spectral fit (Table 2). The luminosity
considered here is that obtained from the hot BB flux as BL is considered to be the major source of the hot BB component. 

Kluzniak (1987) gave the luminosity of the boundary layer as follows, 
\begin{equation}
L_{BL}=(1-\zeta)^{2}\frac{GM\dot{M}}{2R_{NS}}
\end{equation}

Here we consider L$_{BL}$ to be that obtained from the hot BB flux. $\zeta$ is the NS angular velocity in Keplerian units 
($\Omega_{*}$/$\Omega_{k}(R_{NS})$), G is the gravitational constant, M is the mass of the NS, R$_{NS}$ is the NS radius, and $\dot{M}$ 
is the mass accretion rate. The spin frequency of 4U 1705-44 is unknown, hence we make the consideration (similar to D'Ai et al. 2010)
that the difference between the upper and lower kHz QPOs is near the spin frequency for some sources (Méndez \& Belloni 2007; van der Klis 2004).
Based on the study by Ford et al. 1998, this difference would be $\sim$ 330 Hz for 4U 1705-44. Hence $\zeta$ would be $\sim$ 0.15.
Then we estimated $\dot{M}$ and substituted this value in equation 1 to estimate the radius of the BL, which came out to be 14.20 km (A),
15.49 km (B) and 18.14 km (C). D' Ai et al. (2010) estimated a 1.8-2.1 R$_{NS}$ outer radii for the BL based on the spin frequency and
L$_{BL}$/L$_{Disk}$ fraction. This value is not too far from what has been estimated here.

The magnetic field of the NS can also truncate the accretion disk. Alfven radius is proportional to the magnetospheric radius and
based on the transition layer (TL) model, the outer boundary of this layer corresponds to the Alfven radius.  
Kulkarni \& Romanova (2013) obtained a new dependence of magnetospheric radius
on mass accretion rate by performing 3D magnetohydrodynamics simulation of magnetospheric accretion,

\begin{equation} 
r_{m} \sim 2.50 \times 10^{6} \mu^{2/5} \dot{m}^{-1/5} M^{-1/10} R^{3/10} \;cm 
\end{equation}
Here, following Asai et al. (2016), we substituted $\mu$ = B R$^{3}$, with B ranging from 0.5--3.5 $\times$ 10$^{8}$ G, where R=R$_{NS}$=10 km.
In this equation, $\mu$, the magnetic dipole moment in units of 10$^{26}$ G cm$^{3}$, $\dot{m}$ is in units of
10$^{16}$ g s$^{-1}$, M is in units of 1.4 M$_{\odot}$ and R in units of 10$^{6}$ cm. For estimating r$_{m}$, we used the $\dot{m}$
obtained from the spectral fits for A,B and C sections. Substituting these values we estimated, 8-19 km for section A,B and C for magnetic field B = 0.5--3.5 $\times$ 10$^{8}$ G.  

These various radii values estimated using different methods as shown above are well in agreement with each other.

\subsection{Coronal height estimates}
Based on the detected lags, we can estimate the coronal height as per the equation given by Sriram et al. (2019). This equation is derived based on the consideration that the obtained lags are primarily the readjustment timescales of the Corona/sub-keplerian flow.
\begin{equation}
H_{corona}=\Bigg[\frac{t_{lag} \dot{m}}{2 \pi R_{disk} H_{disk} \rho}-R_{disk}\Bigg] \times \beta \; cm
\end{equation}
where H$_{disk}$ = 10$^{8}$ $\alpha^{-1/10}$ $\dot{m}_{16}^{3/20} R_{10}^{9/8} f^{3/20} $ cm, 
$\rho$ = 7 $\times$ 10$^{-8}$ $\alpha^{-7/10}$ $\dot{m}^{11/20}$ $R^{-15/8}$ $f^{11/20}$  g cm$^{-3}$, f = (1-(R$_s$/R)$^{1/2}$)$^{1/4}$ 
(Shakura \& Sunyaev 1973, Sriram et al. 2019).

Here the readjustment velocity in the coronal region v$_{corona}$=$\beta$v$_{disk}$ and $\beta$ is
$\le$ 1 as the coronal viscosity is less than the disk viscosity. Disk 
radius R$_{disk}$ was considered to be 30 km based on the radius obtained from the spectral model (see Table 2)  
and $\beta$ was taken to be 0.1--0.5 based on MHD simulations by Manmoto et al. 1997, Pen et al. 2003 and McKinney et al. 2012. 
Based on these considerations we estimated the coronal height to be 20 km -- 110 km for a lag of 259 s and $\beta$ = 0.1--0.5.
Similarly, if we considered a lag of 38 s, the height estimates are found to be 0.6 km -- 3.37 km.

The values obtained for the coronal height are similar to that obtained for the BL and TL radii, hence we can not rule out the possibility of such 
a layer being responsible for the lags.

\subsection{Disk-corona geometry}

We propose a model similar to the one proposed by Kara et al. (2019) for the Disk Corona geometry of a black-hole transient, with a hot corona
having a static compact core and a geometrically thin accretion disk (see Figure 8a). 
Their model also suggests a varying corona and a stationary disk.

1. In the banana branch the source has a relatively lower hardness, indicating a  low temperature compact corona (Table 3) i. e. a reduced spatial extent of the corona, 
as seen in figure 8a. 

2. As per the spectral modeling the inner disk is stationary along the branch, indicating a constant influx of disk photons as has been
indicated in the figure. This is supported by the fact that the soft light curve in 0.8-3 keV is constant indicating a constant source
of soft photons. 

3. But CCF shows a soft and a hard lag in two different segments, and uncorrelated CCFs in the other segments. Considering a stationary disk, 
this suggests a variation occurring only in the Corona/sub-keplerian flow i. e. source of hard photons. 

4. Lightcurves show a decrease in hard flux when a hard lag was detected, which suggests a decreasing corona. Similarly, the opposite scenario was observed when soft lag was detected. With the observed constant soft flux, we
can fairly conclude that the hard photons could be decreasing with respect to the constant soft photons, thus appearing to produce a hard lag and 
vice versa for soft lag.

5. Segments with no lag suggest that during this time there is possibly not much variation happening in the compact corona/ sub-keplerian flow.

\begin{figure*}
  \centering
  \subcaptionbox{ }[.3\linewidth][c]{%
    \includegraphics[height=3cm,width=.7\linewidth,angle=0]{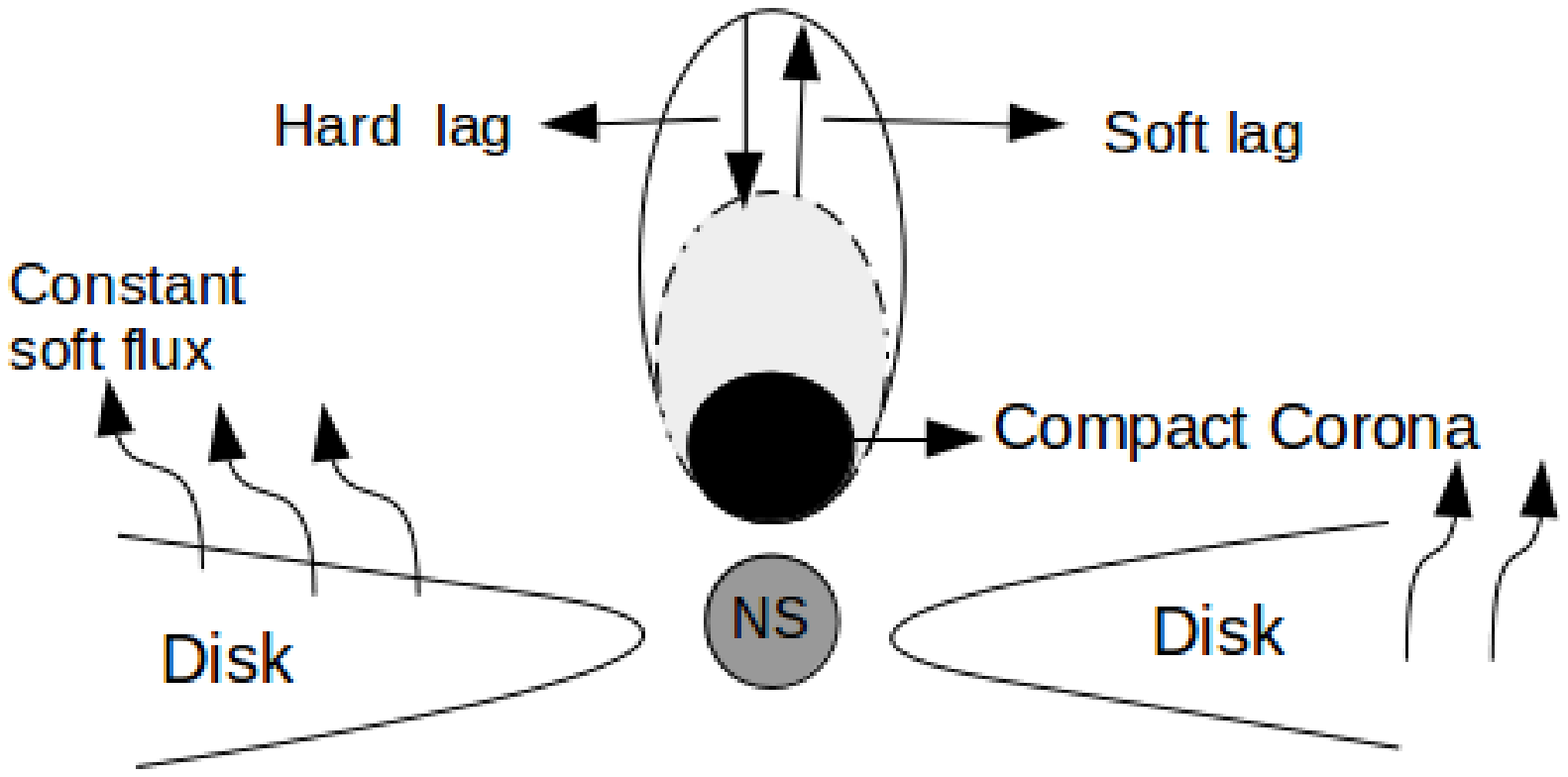}}\quad
  \subcaptionbox{ }[.3\linewidth][c]{%
    \includegraphics[height=3cm,width=.7\linewidth,angle=0]{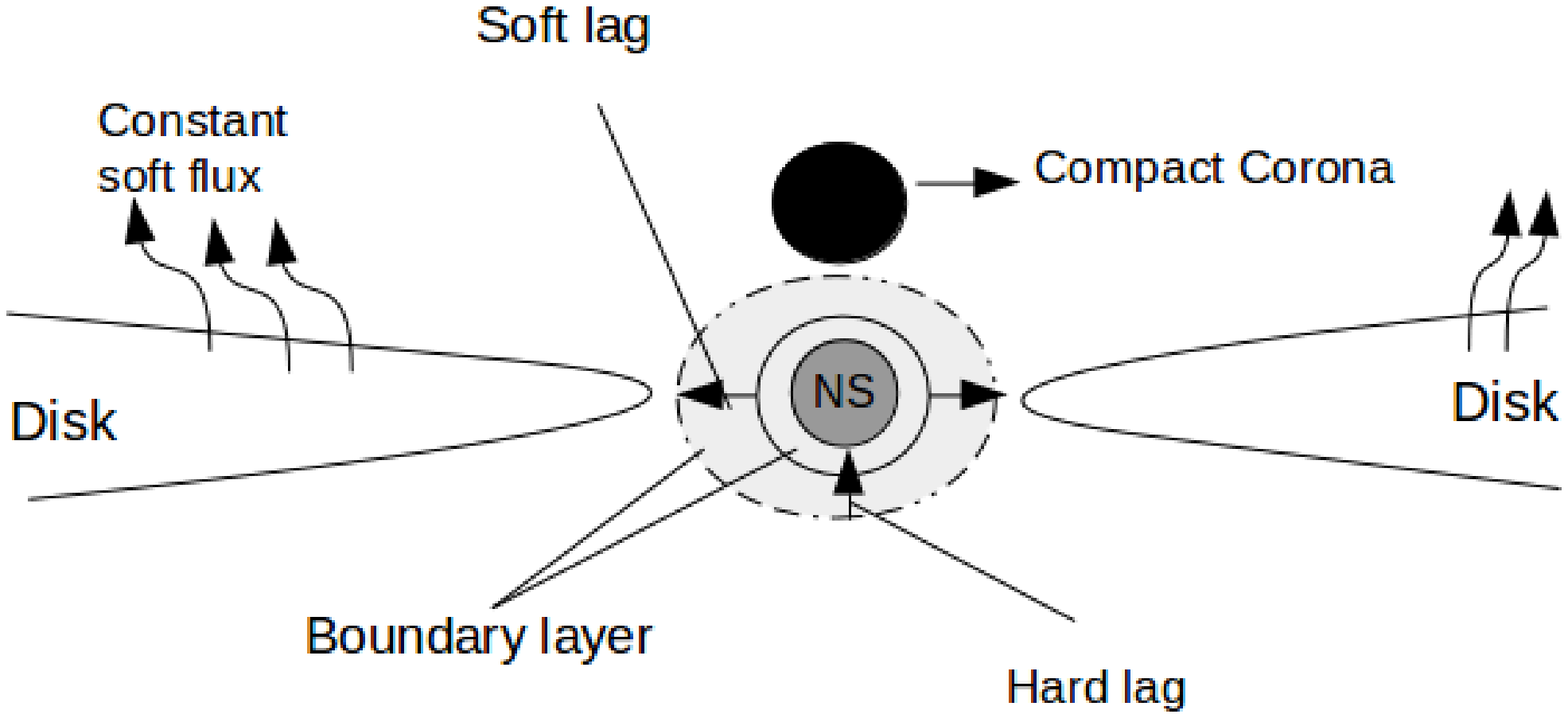}}\quad
  \subcaptionbox{ }[.3\linewidth][c]{%
    \includegraphics[height=3cm,width=.7\linewidth,angle=0]{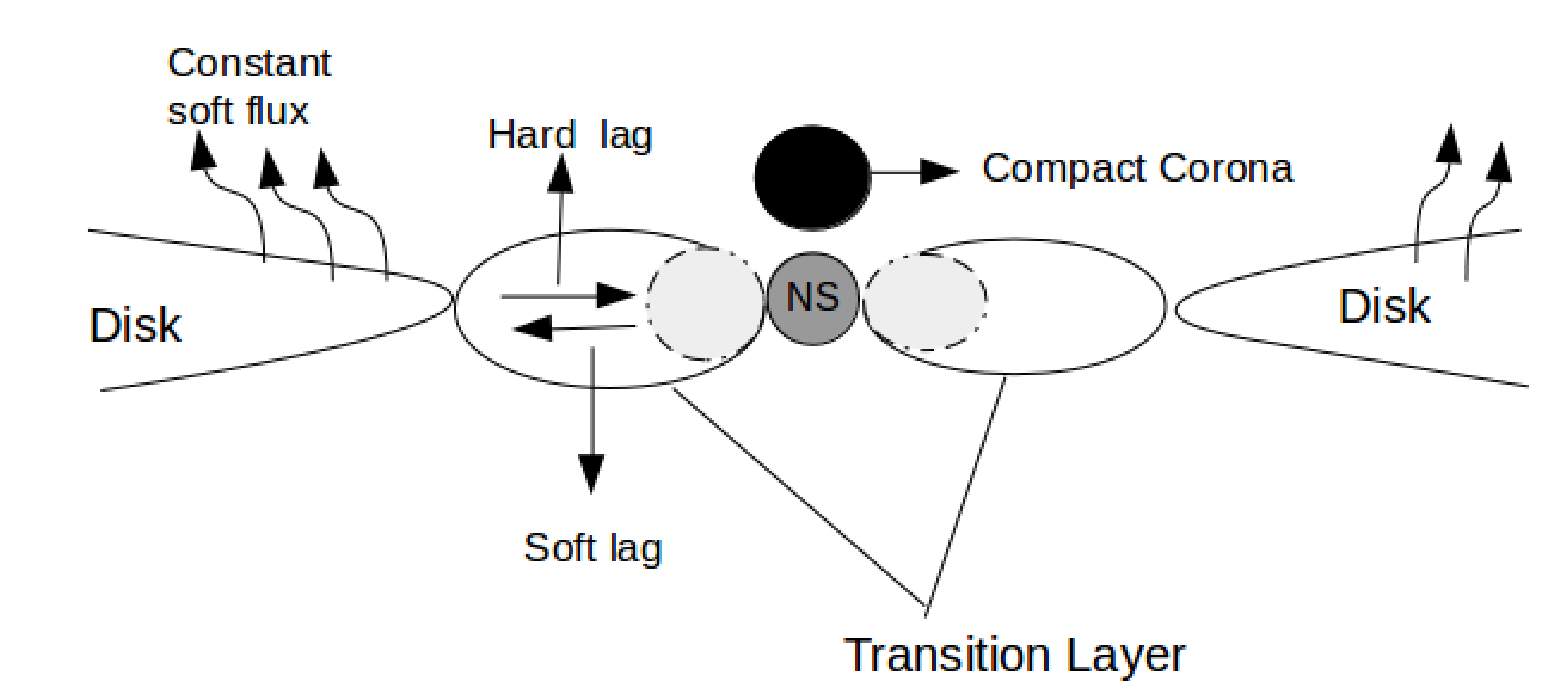}} 
\caption{Proposed geometrical model for accretion disk corona/BL/TL geometry explaining the origin of soft and hard lags (see section 5.3 for a discussion).}
\end{figure*}

Figure 8 summarizes the model in different scenarios where a BL, TL or a compact corona decreasing in size could lead to a hard lag and vice versa for soft lag.
The source of hard photons could be either of the mentioned regions viz. BL, TL, or a compact corona. All scenarios consider constant soft disk photons.
Figure 8a shows a vertically varying coronal structure which shows a hard lag when the height decreases and a soft lag when the height increases. 
Figure 8b shows the scenario wherein the BL around the NS surface is decreasing in its radius giving a hard lag and vice versa for soft lag.
Figure 8c shows a transition layer between the disk and the NS surface which varies in size, giving a hard lag during a size contraction and soft lag during 
size expansion.
Further, thorough observations and timing studies are required to refine this model.

\subsection{Observed lags and a comparison with previous studies}

Lei et al. (2008) performed CCF studies of the Z source Cyg X-2 along the Z track, using
soft (2-5 keV) and hard (16-30 keV) energy bands. They found soft lags less than 200 s and
hard lags less than 400 s. They found more anticorrelated lags in the HB and upper NB,
which led them to associate the anticorrelated lags with a low mass accretion rate.
A similar study performed by Sriram et al. (2012) on the Z source GX 5-1 gave similar results with
soft and hard lags of $\sim$ 31 --920 s, detected mostly in the HB and upper NB. A truncated accretion disk corona geometry was used here to address the observed lags. Ding et al. (2016) found lags of the order of
a few tens of seconds to thousands of seconds in another Z source GX 349+2. Sriram et al. (2019) revealed
similar lags in the HB and NB of Z source GX 17+2 with no definitive trend of lag evolution 
(higher value to lower or vice versa) along the track. Here the disk was found to be near the ISCO. 
This ruled out the possibility of a truncated disk scenario to explain the observed lags. 
The case of atoll sources is not so different when it comes to the estimated lag values.  
Studies were performed on atoll sources 4U 1735-44 (Lei et al. 2013) and 4U 1608--52 (Wang et al. (2014).
and similar lags of a few ten to hundred seconds were detected in both the island and banana states.
For 4U 1705-44, we have used a much softer 0.8-3 keV energy band vs.
10-20 keV energy band, which allows us to study the disk and coronal photons distinctly (similar to Malu et al. 2020).
Here too we have found lags ranging from 38 -- 259 s.

All these studies consistently point towards long term lags in Z and atoll sources, without much concern for their location on the HID. The estimated coronal heights are similar to that of Z source GX 17+2 (Sriram et al. 2019, 2021 and Malu et al. 2020), and based on the obtained lags, 
these estimates would be similar for GX 5-1, Cyg X-2 and GX 349+2. This might point towards a scenario wherein the
accretion disk corona structure might be similar for both these classes of NS LMXBs.
But more observational studies are required to understand this case better.

\section{Conclusion}

For the first time, simultaneous broadband spectral and timing study of the atoll source 4U 1705-44 was performed using AstroSat SXT and LAXPC
data. The source was found to be in the soft banana state during the observations. Results can be summarized as follows.

1. An inner disk radii of 14 Rg was obtained using the full reflection framework spectral modeling, which is well in agreement with previous
studies (Di Salvo et al. 2009, D' Ai et al. 2010, Egron et al. 2013). Inner disk is found to be stationary along the branch from A to C.

2. Disk temperature is similar along the branch, while the hot blackbody component temperature has decreased from 2.21 keV to 2.04 keV along the branch.
A powerlaw index is noted with the presence of a hard tail in the soft state, which contributes 5.85 \% - 7.12 \% of the total flux. 
Powerlaw flux increased along the branch (A to C) from 0.68 - 0.95 $\times$ 10$^{-9}$ ergs cm$^{-2}$ s$^{-1}$. Similarly hot blackbody component flux is found to
be increasing from 3.01-5.91 $\times$ 10$^{-9}$ ergs cm$^{-2}$ s$^{-1}$. Varying hot component fluxes and powerlaw indices indicate towards a varying coronal/sub-keplerian
flow. 

3. Using the luminosity and mass accretion rate obtained from the spectral fits, the boundary layer radius was constrained to be $\sim$ 14-18 km and similarly
using the obtained mass accretion rate, the magnetospheric radius was estimated to be 9-19 km.

4. Energy dependent CCF studies were performed using the 0.8-3 keV soft SXT lightcurves and 10-20 keV hard LAXPC lightcurves. 
An anticorrelated hard lag of $\sim$ 259 s and a correlated soft lag of $\sim$ 38 s were determined from the CCFs. 
This resulted in a height of 0.6-20 km ($\beta$=0.1) for the corona/sub-keplerian flow.

5. We propose a geometrical model for the disk corona/sub-keplerian flow geometry
In our model, the hard lags are produced as a result of a decreasing coronal/BL/TL size and soft lags due to an
increase in the same. The disk would be acting as a source of constant soft flux. These conclusions are
based on the spectral and timing results that indicate varying hard flux and constant Disk flux. 

Similar broadband spectral studies especially are crucial to constrain the accretion disk geometry in NS LMXBs.

\section*{Acknowledgements}
We thank the referee for providing suggestions and comments that have improved the quality of the paper. K.S acknowledges the support from ISRO. The authors sincerely acknowledge the contribution of the LAXPC and SXT instrument teams toward the development of the LAXPC and SXT
instruments onboard the AstroSat. This research has made use of the data collected from the AO cycle 3 of AstroSat observations.
This research work has used the data from the Soft X-ray Telescope (SXT) developed at TIFR,
Mumbai, and the SXT POC at TIFR is acknowledged for verifying and releasing the data via the ISSDC data archive
and also for providing the necessary software tools required for the analysis.
This work also uses data from the LAXPC instruments developed at TIFR, Mumbai and the LAXPC POC at
TIFR is thanked for verifying and releasing the data via the ISSDC data archive. The authors thank
the AstroSat Science Support Cell hosted by IUCAA and TIFR for providing the LAXPC
software which was used for LAXPC data analysis. M.S acknowledges the financial support from the DST-INSPIRE fellowship. S. H acknowledges the support from the CSIR-UGC fellowship.
The author also acknowledges the SERB-CRG for the support.

\section*{Data Availability}
Data used in this work can be accessed through the Indian Space Science Data Center (ISSDC) website 
(https://astrobrowse.issdc.gov.in/astro\_archive/archive/Home.jsp) and is also available with the authors.


\clearpage

\begin{table}
\begin{minipage}[t]{\columnwidth}
\scriptsize
\caption{Best-fit spectral parameters for the 3 sections of the HID which the letters A, B and C respectively represent using the
{\it Diskbb + Bbody + Gaussian(Fe) + Powerlaw} model.
 The subscript BB represents the bbody model and dBB represents Diskbb model.
The flux in units of 10$^{-9}$ ergs cm$^{-2}$ s$^{-1}$ is calculated in the energy band 0.8--70 keV. Errors are quoted at a 90\% confidence level.
Luminosity is in units of 10$^{37}$ erg s$^{-1}.$ assuming the distance 7.4 kpc for 4U 1705-44.}
\label{tab2}
\centering
\begin{tabular}{ccccccccc}
\hline

\hline
\hline
Parameters&A&B&C&\\
\hline
$N_{H}$$\times$ 10$^{22}$ cm$^{-2}$\footnote{Hydrogen column density.}&1.38$_{-0.02}^{+0.02}$&1.38(fixed)&1.38(fixed)\\
$kT_{in}$ (keV)\footnote{Temperature of the Diskbb model.} &1.95$_{-0.08}^{+0.09}$ &2.25$_{-0.08}^{+0.07}$ &2.60$_{-0.03}^{+0.02}$ \\ \\
$N_{dBB}$\footnote{Normalization of the Diskbb model.}& 30.66$_{-3.86}^{+3.92}$&21.06$_{-1.80}^{+2.51}$&12.74$_{-0.81}^{+0.78}$\\ \\

$kT_{BB}$ (keV)\footnote{Temperature of the BB model.} &2.63$_{-0.15}^{+0.21}$ &3.03$_{-0.25}^{+0.35}$ &1.16$_{-0.08}^{+0.09}$ \\ \\
$N_{BB}$\footnote{Normalization of the BB model.}& 0.020$_{-0.005}^{+0.003}$&0.010$_{-0.004}^{+0.005}$&0.010$_{-0.004}^{+0.004}$\\ \\
$E_{Fe}$ (keV) \footnote{Line Energy of the Gaussian model for Iron line.}&6.4 &6.4 &6.4 \\ \\
$\sigma_{Fe}$\footnote{Line width of the Gaussian model for Iron line.}&1.09$_{-0.24}^{+0.23}$ &1.17$_{-0.19}^{+0.17}$&1.15$_{-0.21}^{+0.19}$\\ \\
$EqWidth_{Fe}$\footnote{Equivalent width of the Gaussian model for Iron line in units of keV.}&0.32 $_{-0.2}^{+0.06}$ & 0.39 $_{-0.17}^{+0.05}$ & 0.24 $_{-0.06}^{+0.25}$\\ \\
$N_{Fe}$\footnote{Normalization of the Gaussian model for Iron line.}&0.018$_{-0.007}^{+0.007}$&0.029$_{-0.006}^{+0.006}$&0.035$_{-0.01}^{+0.01}$ \\ \\
$\Gamma_{pl}$\footnote{Power-law index.}&0.40$_{-0.28}^{+0.25}$&0.10$_{-0.14}^{+0.24}$&0.98$_{-0.18}^{+0.18}$ \\ \\
$N_{pl}$\footnote{Normalization of the PL model.}&8.8e-4$_{-0.5e-4}^{+0.5e-4}$ &3.7e-4$_{-2e-4}^{+6e-4}$&9.42e-3$_{-4e-3}^{+8e-3}$\\ \\
Powerlaw flux&0.64&0.78&0.87\\ \\
Total flux&10.32&11.57&13.13\\ \\
Powerlaw fraction & 6.20 \% & 6.74 \% &  6.62 \% \\ \\
F-test probability (powerlaw) & 1.11e-36 & 2.59e-25 & 2.12e-121 \\ \\
F-test probability (Gaussian) & 1.63e-07 & 2.85e-14 & 4.39e-09 \\ \\
L$_{0.8-70 keV}$&6.74&7.55&8.57 \\ \\
$\chi^{2}$/dof&323/352&357/353&382/353\\ \\

\hline
\end{tabular}
\end{minipage}
\end{table}

\clearpage
\begin{table}
\begin{minipage}[t]{\columnwidth}
\scriptsize
\caption{Best-fit spectral parameters for the 3 sections of the HID which the letters A, B and C respectively represent using the 
{\it Diskbb + rdblur*bbrefl + Powerlaw} model. 
The subscript bbrefl represents the bbrefl model, rdblur represents the rdblur model and dBB represents Diskbb model. 
The flux in units of 10$^{-9}$ ergs cm$^{-2}$ s$^{-1}$ is calculated in the energy band 0.8--70 keV. Errors are quoted at a 90\% confidence level.
Luminosity is in units of 10$^{37}$ erg s$^{-1}.$ assuming the distance 7.4 kpc for 4U 1705-44.} 
\label{tab2}
\centering
\begin{tabular}{ccccccccc}
\hline

\hline
\hline
Parameters&A&B&C&\\
\hline
$N_{H}$$\times$ 10$^{22}$ cm$^{-2}$\footnote{Hydrogen column density.}&1.39$_{-0.02}^{+0.02}$&1.39(fixed)&1.39(fixed)\\
$kT_{in}$ (keV)\footnote{Temperature of the Diskbb model.} &1.68$_{-0.02}^{+0.02}$ &1.72$_{-0.11}^{+0.13}$ &1.66$_{-0.03}^{+0.03}$ \\ \\
$N_{dBB}$\footnote{Normalization of the Diskbb model.}& 45.09$_{-2.31}^{+2.47}$&42.49$_{-7.65}^{+8.45}$&45.56$_{-2.92}^{+3.16}$\\ \\
$R_{eff}$\footnote{Effective radius obtained by using Diskbb Normalization, the corrections and the inclination angle of 70$^\circ$}& 10-14 km & 10-13 km & 10-14 km  \\ \\
log$\xi$ \footnote{Ionization parameter.}&3.42 $_{-0.12}^{+0.12}$ & 3.41 $_{-0.13}^{+0.11}$&3.45 $_{-0.11}^{+0.11}$\\ \\
$kT_{bbrefl}$ (keV)\footnote{Temperature of the bbrefl model.} &2.21$_{-0.01}^{+0.01}$ &2.13$_{-0.06}^{+0.08}$ &2.04$_{-0.05}^{+0.05}$ \\ \\
f$_{refl}$\footnote{Reflection fraction.} & 1.63 $_{-0.08}^{+0.08}$& 1.63 $_{-0.06}^{+0.06}$&1.64 $_{-0.05}^{+0.05}$\\ \\
$z$\footnote{Redshift}&0&0&0\\ \\
N$_{bbrefl}$(1 $\times$10$^{-26}$)\footnote{Normalization of the bbrefl model.}& 0.75$_{-0.19}^{+0.26}$&1.04$_{-0.26}^{+.35}$&1.37$_{-0.31}^{+0.36}$\\ \\
$\beta$$_{rdblur}$\footnote{Emissivity Index of the rdblur model.}&-3&-3&-3\\ \\
R$_{in}$(GM/c$^2$)\footnote{Inner disk radii of the rdblur model.}&14.99$_{-3.49}^{+8.02}$&14.45$_{-2.69}^{+5.32}$&13.57$_{-2.52}^{+4.52}$\\ \\
R$_{out}$(GM/c$^2$)\footnote{Outer disk radii of the rdblur model.}&1000&1000&1000\\ \\
i$^{\circ}$$_{rdblur}$\footnote{Inclination angle of the rdblur model.}&70 (fixed) &70 (fixed) &70 (fixed)\\ \\
$\Gamma_{pl}$\footnote{Power-law index.}&0.59$_{-0.21}^{+0.22}$&0.48$_{-0.17}^{+0.16}$&1.07$_{-0.18}^{+0.19}$ \\ \\
$N_{pl}$\footnote{Normalization of the PL model.}&0.002$_{-0.001}^{+0.005}$ &0.002$_{-0.001}^{+0.001}$&0.01$_{-0.006}^{+0.01}$\\ \\
Total flux&11.61&11.62&13.19\\ \\
BB flux &3.01&4.02&5.91\\ \\
DiskBB flux&6.63&6.67&6.23 \\ \\
Powerlaw flux& 0.68&0.85&0.94\\ \\
$R_{sp}$\footnote{Spherization radius}&6.62 km&6.63 km &7.53 km \\ \\
$R_{BL}$\footnote{Boundary Layer radius}&15.74 km&15.74 km&16.52 km \\ \\
L$_{0.8-70 keV}$&7.58&7.58&8.61 \\ \\
$\chi^{2}$/dof&326/353&357/354&395/354\\ \\
\hline
\end{tabular}
\end{minipage}
\end{table}
\clearpage

\begin{table}
\begin{minipage}[t]{\columnwidth}
\scriptsize
\caption{Best-fit spectral parameters for the 3 sections of the HID which the letters A, B and C respectively
 represent using the {\it Nthcomp + Bbody + Gaussian(Fe) + Powerlaw model}.  The subscript bb represents Bbody model.
The flux in units of 10$^{-9}$ ergs cm$^{-2}$ s$^{-1}$ is calculated in the energy band 0.8--70 keV. Errors are quoted at a 90\% confidence level.
Luminosity is in units of 10$^{37}$ erg s$^{-1}$, assuming the distance 7.4 kpc for 4U 1705-44.} 
\label{tab1}
\centering
\begin{tabular}{ccccccccc}
\hline

\hline
\hline
Parameters&A&B&C&\\
\hline
$N_{H}$$\times$ 10$^{22}$ cm$^{-2}$\footnote{Hydrogen column density.}&1.47$_{-0.11}^{+0.11}$&1.47(fixed)&1.47(fixed)\\ \\
$\Gamma_{Nthcomp}$\footnote{Nthcomp Power-law index.} & 2.48 $_{-0.13}^{+0.15}$& 2.56 $_{-0.14}^{+0.15}$ & 2.41 $_{-0.14}^{+0.16}$\\ \\
$kT_{e}$ (keV)\footnote{Electron temperature (Nthcomp).} & 2.92 $_{-0.13}^{+0.17}$ & 2.96$_{-0.18}^{+0.22}$& 2.79$_{-0.16}^{+0.20}$  \\ \\
$kT_{bb}$ (keV)\footnote{Seed photon temperature (Nthcomp).} & 0.96 $_{-0.06}^{+0.06}$& 1.05 $_{-0.04}^{+0.04}$ & 1.07 $_{-0.05}^{+0.05}$ \\ \\
$N_{Nthcomp}$\footnote{Normalization of the Nthcomp model.}& 0.32 $_{-0.03}^{+0.03}$ & 0.31 $_{-0.02}^{+0.02}$& 0.33 $_{-0.02}^{+0.02}$ \\ \\
$kT_{BB}$ (keV)\footnote{Temperature of the BB model.} &0.34$_{-0.03}^{+0.03}$ &0.36$_{-0.02}^{+0.02}$ &0.36$_{-0.02}^{+0.02}$ \\ \\
$N_{BB}$\footnote{Normalization of the BB model.}& 0.021$_{-0.004}^{+0.005}$&0.022$_{-0.001}^{+0.001}$&0.022$_{-0.001}^{+0.001}$\\ \\
$E_{Fe}$ keV \footnote{Line Energy of the Gaussian model for Iron line.}&6.4 &6.4 &6.4 \\ \\
$\sigma_{Fe}$\footnote{Line width of the Gaussian model for Iron line.}&0.99$_{-0.72}^{+0.46}$ &0.99$_{-0.45}^{+0.47}$&1.02$_{-0.32}^{+0.38}$\\ \\
$N_{Fe}$\footnote{Normalization of the Gaussian model for Iron line.}&0.007$_{-0.006}^{+0.006}$&0.008$_{-0.007}^{+0.007}$&0.015$_{-0.009}^{+0.009}$ \\ \\
$\Gamma_{pl}$\footnote{Power-law index.}&0.36$_{-0.27}^{+0.25}$&0.13$_{-0.22}^{+0.20}$&0.63$_{-0.28}^{+0.26}$ \\ \\
$N_{pl}$\footnote{Normalization of the PL model.}&7.5e-4$_{-0.5e-4}^{+1e-3}$ &3.9e-4$_{-2e-4}^{+5e-4}$&2.38e-3$_{-1e-3}^{+4e-3}$\\ \\
Total flux & 10.59 & 11.86&13.45\\
L$_{0.8-70 keV}$&6.91&7.74&8.78 \\ \\
$\chi^{2}$/dof&274/350&270/351&327/351\\ \\
\hline
\end{tabular}
\end{minipage}
\end{table}

\nocite{*}
\bibliography{ref}{}
 \bibliographystyle{mnras}


\bsp	
\label{lastpage}
\end{document}